\documentclass[twocolumn,showpacs,preprintnumbers,amsmath,amssymb]{revtex4}
\usepackage{graphicx}% Include figure files
\usepackage{dcolumn}% Align table columns on decimal point
\usepackage{bm}% bold math
\usepackage[latin1]{inputenc}  % accents 8 bits dans le source
\usepackage[T1]{fontenc}       % accents dans le DVI
\usepackage{gensymb}

\begin{document}

\preprint{}

\title{Ultrasonic probing of the elastic properties of PMMA bead packings and their rearrangement during pressure sintering}
\author{Vincent Langlois}
 \email{Vincent.Langlois@univ-mlv.fr}
\author{Xiaoping Jia}
 \email{Xiaoping.Jia@univ-mlv.fr}
\affiliation{Université Paris-Est, Laboratoire de Physique des
Matériaux Divisés
et des Interfaces, CNRS UMR 8108, Cité Descartes, 77457 Marne-la-Vallée, France }

\date{15 october 2009}

\begin{abstract}
Ultrasound transmission in PMMA spherical bead packings is investigated during the sintering process under stress. Velocity and amplitude measurements of coherent longitudinal waves are performed to monitor the evolution of the elastic properties of the solid frame from noncohesive packing to sintered granular material. Comparison between the experimental velocity data and the prediction by a contact model [Digby, J. Appl. Mech. 48, 803, (1981)] reveals the crucial role of the bonding effect on the mechanical behavior of granular compacts. By using the correlation technique of acoustic speckles, we also observe the important rearrangements in granular packings before the onset of sintering.
\end{abstract}

%\keywords{Ultrasonic measurements, Sintering and bonding, Granular materials, Rearrangement of particles, Correlation of acoustic speckles.}

\maketitle

\section{Introduction}

The consolidation of powders is an important manufacturing process to fabricate the materials and parts from metallic, ceramic and polymeric powders \cite{rahaman05}. In general, production of powder parts consists of two basic steps. First, powder is consolidated into the desired shapes via isostatic or uniaxial (die) pressing to form a porous powder compacts referred to as a green body. Second, the formed powder compact is sintered at high temperature. Sintering bakes out lubricant if any from the green body, promotes further densification and, most importantly, strengthens the part by soldering or bonding particles together by atomic diffusion in the solid state. The driving force for solid state sintering is basically attributed to the reduction in free surface energy of the consolidated particles \cite{rahaman05,artz82}. This reduction in energy can be accomplished either by densification (shrinkage) of the body with mass transport from the particles into the pores or by grain coarsening without actual decrease in the pore volume. Three stages are usually distinguished. In the initial stage the contact area between particles increases accompanied by the neck growth. During the intermediate stage considerable densification occurs before isolation of the pores. The final stage involves densification from the isolated pore state to the final densification \cite{rahaman05,artz82}. 

Numerous simplifying models have been developed for analyzing the densification process and evaluating the effects of various processing parameters (temperature, sintering time, stress) \cite{artz82,helle85,larsson96}. For the initial stage, the two-sphere model allows to characterize mechanisms of sintering such as diffusional redistribution of matter, plastic deformation or creep flow at contact. This model describes adequately the densification of a powder compact via the neck growth between contacting particles and the increase in the coordination number \cite{fischemeister78, swinkels83}. From the experimental point of view, the commonly used techniques for in-situ monitoring of sintering are the dilatometric measurements \cite{rahaman05}. These techniques are however not sensitive to sintering mechanisms which do not involve the shrinkage of powder compacts. Specific surface area which is sensitive to the effects of all mass transport mechanisms has been shown as a useful parameter for evaluating the sintering kinetics. However, the standard gas adsorption technique for measurement of specific surface area requires the sample to be cooled and precludes in-situ measurements \cite{rahaman05,martin98}.

Ultrasonic inspection offers non intrusive in-situ measurements of porous granular materials and has been used for monitoring the sintering process \cite{martin98,dawson98}. The acoustic properties of these porous materials are in general characterized by the properties of the contact network forming the solid frame, the pore fluid if any, and frame-fluid interactions \cite{digby81,makse04}. In the long wavelength limit, the effective elastic moduli of the granular solid critically depend on the microstructure of the material, including the particle shape and size distribution, interparticle interaction and stiffness, coordination number and packing structure \cite{digby81,winkler83,makse04,garcia99}. These features change over the course of sintering; however the effects of these changes on sound propagation are not well understood. Recently, it was reported that the measured increase in ultrasonic velocity during the initial stage of sintering may be correlated with the reduction of specific surface area in which no significant shrinkage is involved \cite{martin98}. A simplifying model based on the pure Hertz contact between spherical particles has confirmed such an available relationship between the acoustic velocity increase and the reduction of specific surface area during the early stages of solid-state sintering \cite{garcia99}. However, the influence of the particle bonding on the acoustic velocity during sintering remains unclear.

In this work, we present new ultrasonic measurements conducted in model granular packings of PMMA beads under stress together with density measurements during the sintering process. Both coherent longitudinal wave and high-frequency scattered waves \cite{jia99}  are investigated for monitoring the evolution of the properties of the contact networks from green (non cohesive) compacts to sintered (cohesive) ones during the thermal cycle. Measured acoustic velocities are compared with the prediction of the theoretical model developed by Digby \cite{digby81}  which accounts for the effect of bonding at the interparticle contact. On the other hand, the transmitted amplitude of the coherent longitudinal wave reveals much more important change than that of the velocity near the glass transition of the PMMA material.

Furthermore, we make use of the correlation technique of multiply scattered ultrasound to investigate the rearrangements of grains by the thermal cycling during the green body state which has showed an important effect on the compaction of granular materials \cite{chen06,vargas07,divoux08}. Rearrangement which is not accounted in analytical models of sintering \cite{martin03,luding05} could however have the influences on the mechanical properties of sintered materials \cite{rahaman05} .

\section{Experiments}

\subsection{Experimental set-up}

Our porous granular material is formed by sintering the PMMA spherical beads of $2R \approx 630-800~{\mu}m$. The random granular packings are prepared by a rain deposition in an oedometric cell (i.e., a die) of diameter of 32 mm, made of duralumin. They are compacted by one cycle of loading-unloading corresponding to uniaxial stress $P = 100-500~kPa$; the thickness of the samples is $h_0 \approx  12~mm$ with solid volume fraction of $\phi_s \approx 0.61$. The variation of the thickness $\Delta h$ is measured by a displacement sensor in which the length variation of the delay line is accounted. These granular materials under constant stress $P$ are then heated by external resistor bands up to a temperature close to the glass transition $T_g\ (\approx 105\degree C)$ of the PMMA (Fig 1). The temperature is controlled by a thermocouple $(T)$ situated at the bottom of the cell. The heating schedule consists of three steps : the first is the increasing of temperature at a constant rate of 1.5°C/min; the second is isothermal dwells at three temperatures: 90°C, 95°C or 100°C close to $T_g$ for 30~minutes; and the third is the cooling at a rate of 1.5°C/min (see below).

A large ultrasonic transducer of diameter 30 mm with a delay line is put at the top of the bead packing. Four-cycle tone burst excitations of $1.3 {\mu}s$ duration centered at a frequency of 300 kHz are applied to this longitudinal source transducer. A small ultrasonic receiver $(D)$ of diameter 1.5~mm is used to enhance the detection of scattered ultrasonic waves \cite{jia99}, placed at the bottom of the cell in direct contact with the beads (Fig. 1). The signal-to-noise ratio is improved by repetitive averaging of the detected signals using digital oscilloscope and sent to microcomputer for further signal processing.

\begin{figure}[]
  \includegraphics[width=5cm]{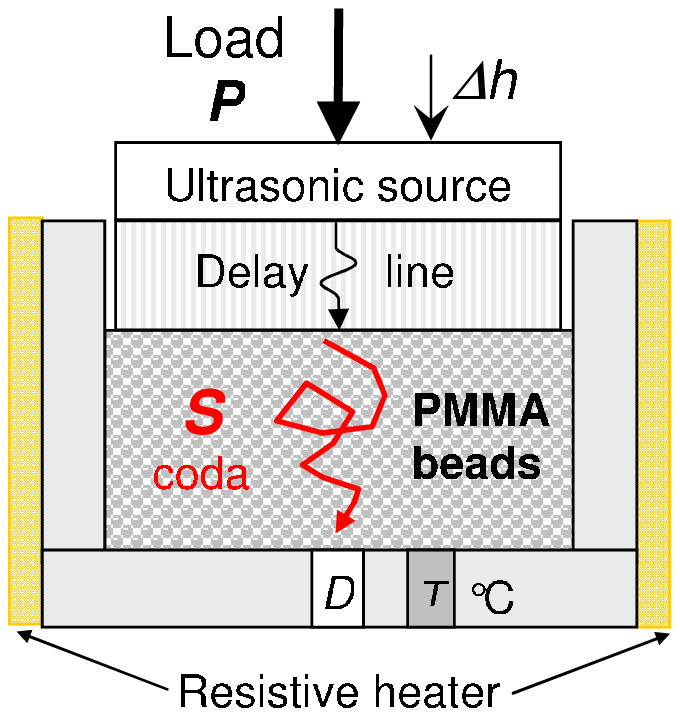}
  \caption{\label{fig:setup}Sintering setup coupled with ultrasonic measurements. Ultrasound propagating in the bead packing is detected by a small transducer $(D)$.}
\end{figure}
\begin{figure}[]
  \includegraphics*[width=7cm]{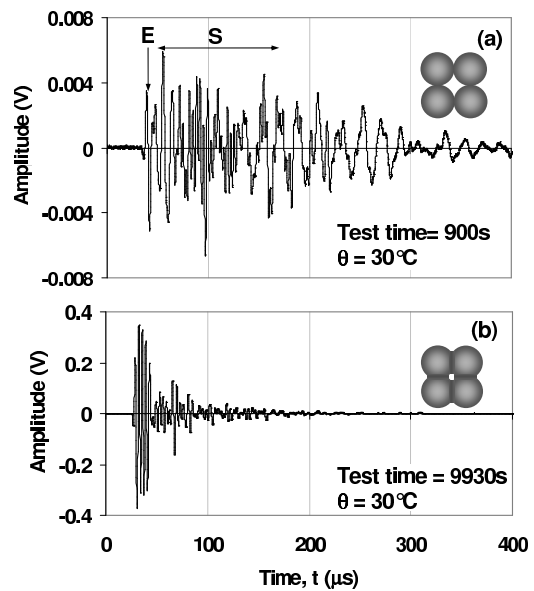}%includegraphics* * permet de découper la figure
  \caption{\label{fig:fig2}Ultrasonic transmission through a PMMA bead packing measured at room temperature (30°C) before (a) and after (b) sintering.}
\end{figure}

\subsection{Transitional behavior of ultrasonic wave transport} %2.2

Fig. 2a shows a typical ultrasound transmission in a non cohesive granular packing. As described previously [13], the transmitted signal is basically composed of a primary coherent component and a strongly fluctuating incoherent component. It has been shown by the spectral analysis that the low-frequency coherent pulse $E \ (\approx 50~kHz)$ at the leading edge of the transmitted signal corresponds to a self-averaging effective longitudinal wave propagating ballistically, while the high-frequency incoherent signal $S \ (\approx 250~kHz)$ arriving mostly at later times is attributed to specklelike scattered waves by the inhomogeneous distribution in the contact networks. The effective wave velocity is measured by the time-of-flight of the coherent $E$ pulse, being about $V_L \approx 700~m/s$. The associated wavelength is to be about $\lambda_E \approx 10~mm$ for coherent $E$ wave much larger than the grain size 2R, and $\lambda_S \approx 2~mm$ for specklelike scattered $S$ waves close to the grain size. The significant difference between $E$ and $S$ waves spectra indicates that an adequate spectral filtering of the total transmitted signal permits to separate the respective contribution of coherent and scattered waves. Note that the configuration-specific acoustic speckles are very sensitive to any change in the microstructure such as rearrangements.

During the sintering process, both the contact area and the strength of bonding between the particles are expected to increase with test time which shall lead to an important evolution in the elastic properties of the granular solid from non cohesive state to sintered one. Fig. 2b displays the ultrasonic transmission through a sintered porous compact cooled down to room temperature $(\approx 30\degree C)$. The clear difference in the waveforms transmitting through the bead packings before and after sintering illustrates a significant change in the wave transport: coherent propagation becomes dominant whereas inherent scattered waves are considerably reduced. As described below, the acoustic velocity increases progressively during sintering and it is accompanied by a considerable enhancement in its transmission amplitude of the coherent wave.

\subsection{Acoustic velocity measurements versus densification}%2.3

Fig. 3a shows both the evolution of the longitudinal acoustic velocity $V_L$ and the variation of the thickness (density) of the sample $\Delta h$ under constant stress $P \approx 500~kPa$ during the heating cycle. The solid volume fraction $\phi$ or relative density of the sample is monitored via $\rho = M/(\pi D^2 h)$ with M the mass of the sample. Distinct responses are observed at three different steps of the thermal cycle. At the step I where the temperature is below the glass transition $T_g\ (\approx 105\degree C)$, the medium remains as a non cohesive granular packing and the acoustic velocity does not evolve significantly. At the step II when the temperature increases and approaches to the glass transition, a significant densification or shrinkage of the granular packing $\Delta h~(=~h_0 - h)$ is observed. The densification of the medium coincides with the important increase of the acoustic velocity. At the step III the sintered bead packing is cooled down and becomes a porous material. The continuous increase observed in the velocity at this cooling step where the solid skeleton is almost frozen is likely associated with the increase of the material modulus of the PMMA \cite{asay69} .

To investigate the relationship between acoustic velocity and shrinkage measurements, we plot in Fig. 3b the evolution of $V_L$ versus $\Delta h$. A clear correlation between these two parameters is observed, especially during the step II of heating where the sintering process is expected to set on. Such a correlation between acoustic velocity and densification is found for various experimental conditions of sintering process, as shown in Fig. 4. Note however that for the low stress $P \approx 250~kPa$, there exists a significant velocity increase $(\approx 10\%)$ at the very initial stage of sintering, but without measurable densification (inset of Fig. 4). Such an increase in the velocity could be associated with the change in the contact stiffness caused by the bonding between PMMA beads via molecular diffusion (see § 3.1).

\begin{figure}[]
  \includegraphics*[width=7cm]{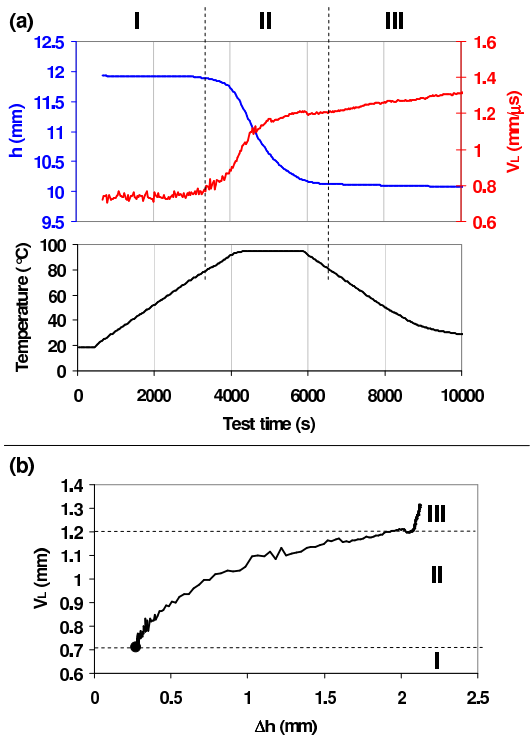}
  \caption{\label{fig:fig3}(a) Evolution of the longitudinal acoustic velocity $V_L$ and the sample thickness h during the heating cycle at three different ranges of temperature under $P \approx 500~kPa$. (b) Acoustic velocity $V_L$ versus shrinkage $\Delta h/h_0.$}
\end{figure}

\begin{figure}[]
  \includegraphics*[width=7cm]{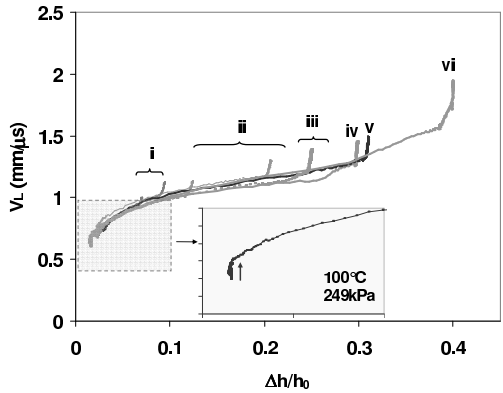}
  \caption{\label{fig:fig4} Acoustic velocity $V_L$ versus shrinkage $\Delta h/h_0$ for different temperature plateaus and strain. (i:~90°C~500kPa~- ii:~95°C~500kPa~- iii:~100°C~250kPa~- iv:~100°C~500kPa~- v:~100°C~750kPa~- vi:~120°C~500kPa.)}
\end{figure}

\subsection{Amplitude measurements of the coherent wave transmission}%2.4

As described above, the coherent wave becomes progressively dominant in the total ultrasound transmission as the sintering sets-up. To evaluate more quantitatively this evolution, we examine the spectrum of the transmitted coherent wave obtained at different steps of the thermal cycle. More precisely, the spectrum is calculated with the first oscillation at the leading edge, windowed out of the coherent pulse $E$. Fig. 5a illustrates the corresponding spectra of coherent longitudinal waves, in which we observe an important increase in wave amplitude and also a significant shifting of dominant transmission towards high-frequency components. For comparison, we present in Fig. 5b, the evolution of both the acoustic velocity and the spectral amplitude measured at 300 kHz in the course of the heating cycle. It appears clearly that the change in the amplitude of coherent wave transmission is much more important than that in the wave velocity particularly at the initial stage of sintering. The amplitude decrease observed at the cooling step is due to the effects of thermal expansion causing the contact loss between the ultrasonic transducers and the sample.

\begin{figure}[]
  \includegraphics*[width=7cm]{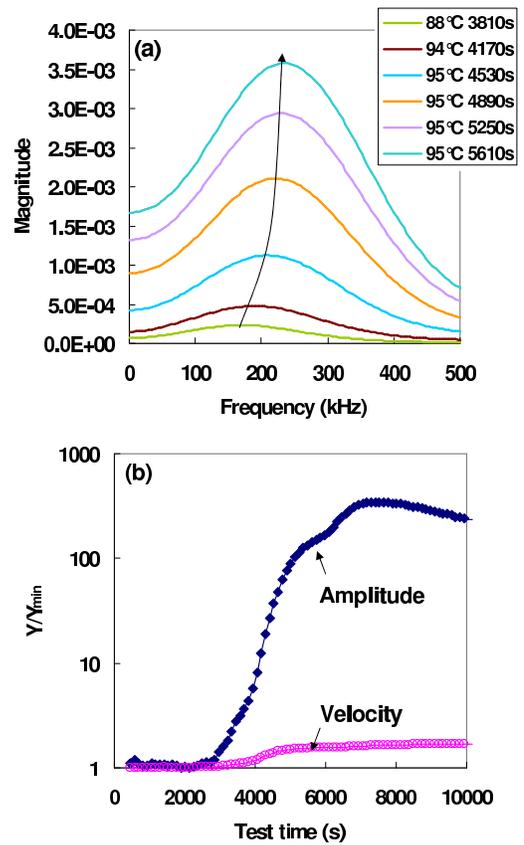}
  \caption{\label{fig:fig5} (a) Spectrum evolution of the coherent longitudinal wave. (b) Relative variations $Y/Y_{min}$ of the transmitted amplitude centred at $300~kHz$ and the acoustic velocity $V_L$.}
\end{figure}

The evolution of the coherent wave transmission may be understood by the microstructure's change of the medium. Previous works showed \cite{dawson98,brunet08} that the attenuation mechanism of high-frequency coherent waves in porous granular materials is dominated by wave scattering. The damping can be characterized by the attenuation coefficient $\alpha$, of the coherent wave displacement field. The scattering attenuation strongly depends on the ratio $\lambda_E/d^*$ of the wavelength to the characteristic size of heterogeneity $d^* (= d)$; for the Rayleigh scattering one has $\alpha = (\lambda_E/d^*)^4 = (fd^*/V_L)^4$. With such a picture in mind, the considerable increase of the transmission amplitude may be attributed to the decrease of scattering attenuation during the sintering process, associated with the increase in acoustic velocity $V_L$ or the decrease in heterogeneity size $d^*$. Our experimental finding shows in addition to the acoustic velocity, the amplitude measurement may provide the useful information on smaller length scales of the porous compacts.

\section{Comparison with the effective medium theory}%3

To interpret the acoustic velocity data, we compare with the predictions by the effective medium theory (EMT) based on the adequate local contact force law \cite{digby81,winkler83,berryman95,makse04,garcia99}. Within the framework of this mean field approach, the kinematics of particles is imposed by a homogeneous macroscopic strain field in which there is no local rearrangement (i.e., affine approximation). From the assumption that both the distributions of contact number and contact orientation are isotropic, these models allow to predict the compaction and the elastic modulus of the packing as a function of the isostatic pressure.

\subsection{Contact stiffness and acoustic velocity}%3.1

Unlike the previous work where the pure Hertz contact model was considered \cite{garcia99}, we describe here the Digby contact model which may account for the effect of cohesion on acoustic velocities from green compacts to sintered one \cite{digby81} . This model consisted in modelling the consolidated rock by a random packing of identical bonded elastic spheres of radius $R$. As shown in Fig. 6a, the two spherical particles in contact are assumed to be initially bonded over a plane circular contact area of radius $b$, resulting from cementation or sintering. As isostatic stress applied on the packing increases, the total area of contact increases to radius $a~(> b)$. The radius $b$ is independent of the applied stress; $b = 0$ corresponds to the case of noncohesive particles. By analogy with the Hertz theory, the normal stiffness is given by $D_n = 4 \mu a/(1-\nu)$ ($\nu$  is the Poisson ratio of the grain material), while an analogy with the Mindlin's model gives the tangential contact stiffness $D_t = 8 \mu b/(2-\nu)$ if the annular contact region between radii $a$ and $b$ is assumed to perfectly sliding. By a statistical analysis, the effective elastic Lamé constants $\lambda$ and $\mu$ can be derived from the contact stiffnesses $D_n$ and $D_t$ or from the contact areas $a$ and $b$. Accordingly, the longitudinal and transverse acoustic velocities are given by,
\begin{eqnarray}
{V_L}^2=\frac{\mu Z}{5 \pi \rho}\left[ \frac{3 \overline{a}}{1-\nu}+\frac{4 \overline{b}}{2-\nu}\right] \\
{V_T}^2=\frac{\mu Z}{5 \pi \rho}\left[ \frac{\overline{a}}{1-\nu}+\frac{3 \overline{b}}{2-\nu}\right] 
\end{eqnarray}
where $\rho$ is the particle density, $Z$ the coordination number,   and  $\overline{b}=b/R$ and $\overline{a}=a/R (\ll 1)$ the dimensionless areas.

\begin{figure}[]
  \includegraphics*[width=7.5cm]{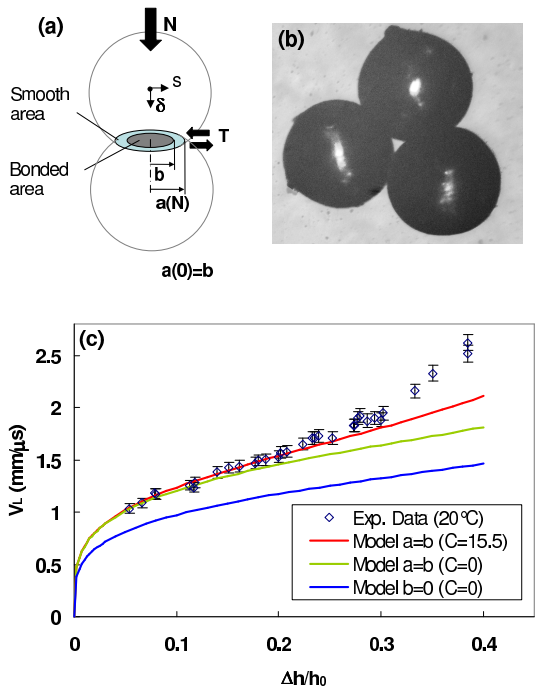}
  \caption{\label{fig:fig6} (a) Contact geometry between two spheres in the Digby model ; (b) Optical observation of the sintered pore shape by microscopy ; (c) Comparison between theoretical predictions and experimental data of longitudinal wave velocity, as a function of the shrinkage.}
\end{figure}

Here, we propose to model the effect of sintering on acoustic velocities via a transformation of the nature of the interparticle contact. More precisely, before the onset of sintering, the contact area a is purely hertzian with $b = 0$, and after sintering, the total area of contact a is bonded with $b = a$. The change of contact may be realized with different mechanisms of material diffusion at the contact \cite{rahaman05} and becomes especially efficient at the glass transition temperature of the PMMA. During further growth of contact area (i.e., neck) with sintering, the assumption of bonded contact, i.e. $b = a$, is still applied. Compared to the acoustic velocity calculated without considering the cohesion effect \cite{garcia99}, namely keeping $b = 0$ in eqs. (1) and (2), the picture that we propose for sintering predicts an important increase of acoustic velocity, up to $25\% $ for the longitudinal one $V_L$ in the PMMA compacts ($\nu = 0.33$). Moreover, this picture suggests that the increase of the acoustic velocity due to the change of contact stiffness may not be associated with the shrinkage or increase of contact area, as observed experimentally at the beginning of sintering process (inset of Fig. 4).

\subsection{Coordination number and contact area}%3.2

For evaluating longitudinal wave velocities (eq. 1), one also needs the information concerning the evolution of the coordination number $Z$ and contact area $b$ during sintering or/and shrinkage. Within the approximation of homogeneous strain field, many authors have used the radial distribution function to relate the local overlap $\delta$ between two spheres, the contact radius $a$, and the average coordination number $Z$ to the macroscopic solid volume fraction $\phi$ of a random packing subject to densification \cite{artz82,helle85,larsson96}. A simple relation was derived for the increase in coordination due to the reduction of particle distances:
\begin{equation}
Z = Z_0 + C (\phi-\phi_0)
\end{equation}
where $Z_0\ (\approx 7.3)$ and $\phi_0\ (\approx 0.64)$ are respectively the initial coordination number and solid volume fraction of the packing, and $C\ (\approx 15.5)$ accounts for the recruitment of contacts.

During sintering, new contacts may also form as a result of the change in contact geometry. The neck shape depending on the mechanisms involved: diffusion gives pore rounding, while plastic flow and creep give cusped pores. Previous microscopic observations by Swinkels et al \cite{swinkels83}  on PMMA powders of $100 {\mu}m$ sintered at 110°C and 700 kPa showed that as the necks grow, the free surfaces of the particles expand outwards into the pore space maintaining a sharply cusped pore shape even at high values of $\phi\ (\approx 0.9)$. Such contact geometry resembles that after pressure densification with little contribution from diffusion. With optical microscopy, we have also investigated the pore shape of our sintered PMMA compacts. As shown in Fig. 6b the pore shape appears rather cusped than rounded, similar to those previous observations. During densification, both the average coordination number increases steadily and the existing contacts grow in area. Artz \cite{artz82} analyzed the evolution of both processes by considering a concentric increasing of the effective particle radius in an average Voronoi cell and derived the average contact area $ \pi \langle b^2 \rangle \ or \ \pi \langle a^2 \rangle$ by averaging over all existing contacts: $Z_0$ contacts of maximum size (initial contacts) and some progressively smaller ones (newly formed). Following the same approach, we may determine the average radius of contact $\langle b \rangle$ or the product $\langle Z \overline{b} \rangle $ by
\begin{equation}
\overline{b}=\frac{Z_0}{R"} \sqrt{R"^2-1} +C \int_{r=1}^{r=R"} \sqrt{R"^2-r^2}dr
\end{equation}
with $R"$ the dimensionless effective particle radius \cite{artz82}. Eq. 3 can be analytically expressed in terms of the overlap $\delta$ (not shown here). From the approximation of homogeneous strain field, the average overlap between two spheres can be related to the uniaxial deformation by $\delta/R = \Delta h/3h_0$. Note that for an isotropic deformation, one would have $ \delta/R = \Delta h/h_0$.
Fig. 6c presents the longitudinal velocities $V_L$ as a function of the shrinkage $\Delta h/h_0$, calculated from eqs. 1, 3 and 4 with $\mu = 2.04 GPa$. The effects of contact bonding and recruitment on the velocity are shown by non vanishing value of $b\ (= a)$ and $C$, respectively. For comparison, the acoustic velocities of the longitudinal wave measured in different sintered PMMA compacts at room temperature are also shown there. Measurements conducted at room temperature may simplify the comparison with the model which needs the information about the elastic modulus and the density as a function of temperature. Fig 6c shows a fairly good agreement between velocity data and theoretical predictions ($b = a$ and $C = 0$), especially for low shrinkage $\Delta h/h_0$. More quantitatively, the effect of newly formed contacts is $15\%$ at most, while the strength enhancement of contact by the bonding effect plays a crucial role. The discrepancy observed at important shrinkage (density) arises probably from the applicability of the Digby model valid for $b \ll R$ and from the approximation of an isotropic deformation. 

\section{Rearrangements of the contact networks}

As mentioned earlier, most of analytical models for sintering adopt the assumption of homogeneous strain field and neglect the local rearrangement during densification. It is shown however that under external stress, noncohesive granular packings (i.e., green compacts) develop a highly inhomogeneous contact network of stressed particles. Recent works showed the important effects of thermal cycling on the densification of granular material \cite{chen06,vargas07,divoux08}. It is reported that either the difference between the thermal dilatation of the container and the grains, or the locally heterogeneous dilatation of the grains due to the thermal gradient may be the cause of the rearrangements leading to the densification. Such phenomena could also increase the coordination number and affect the contact area during heating cycle, and modify the radial distribution function used for describing the evolution of contacts during densification \cite{vargas07,martin03,luding05} . It is also now recognized that inhomogeneities in green compacts can seriously hinder the ability to achieve high density and adequately control the fabricated microstructure in sintered powder materials\cite{rahaman05}.

\begin{figure}[h]%htbp
  \includegraphics*[width=7cm]{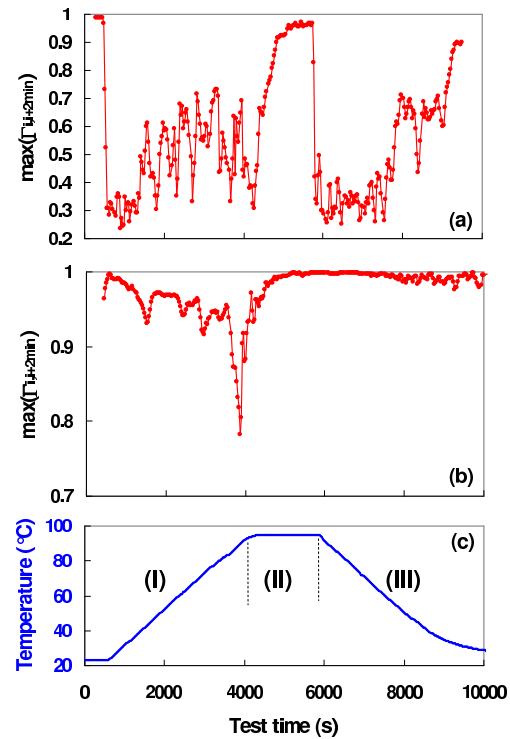}
  \caption{\label{fig:fig7} Cross-correlation function of high-frequency acoustic speckles: (a) in glass bead packings with frequency width of 400-600 kHz; (b) in PMMA beads with frequency width of 225-350 kHz.
}
\end{figure}

Here we investigate the dynamics of the network rearrangement during thermal cycling by using the configuration-specific acoustic speckles. To this end, we define a degree of resemblance $\Gamma_{i,j}$ between two speckles signals $S_i (t)$ and $S_j (t)$ recorded at an interval of 2 minutes during the thermal cycle \cite{jia01}:
\begin{equation}
\Gamma_{i,j} (\tau)= \frac{C_{i,j} (\tau)} {\sqrt{C_{i,i}(0) C_{j,j}(0)}}
\end{equation}
Here $C_{i,j} (t)$  is the cross-correlation function given by $C_{i,j} (t)= \int S_i(t).S_j(t-\tau)d\tau$ with t, the lag time. The maximum value of  $\Gamma_{i,j}$, $max[\Gamma_{i,j}]$, characterizes their resemblance or correlation. It is found [24] that  $\Gamma_{i,j}$ allows to monitor the evolution of the contact network, resulting either from the relative motion of grains or from the creep at the level of contact \cite{jia09}. When the PMMA bead packing is heated up close to the glass transition temperature, two phenomena are expected: the rearrangement and the neck growth between particles in contact. In order to evaluate the effect of each phenomenon, respectively, we perform the similar measurement with the glass bead packings in which no sintering occurs at the range of temperature used in this work.

Fig. 7 shows the evolution of the maximum value of  $\Gamma_{i,j}$ during the heating cycle. For glass bead packings, the important decorrelation appears in the heating step I and cooling step III, while $max[\Gamma_{i,j}]$ approaches to unity during the isothermal dwell implying a slowdown in the microstructural evolution. These experimental findings are consistent with previously reported the granular compaction induced by the temperature change via rearrangements of particles \cite{chen06,divoux08}. For PMMA bead packings, the significant decorrelation is observed at the step I of the thermal cycle and the beginning of step II, associated with the relevant rearrangement in the green or noncohesive state of bead packings. The magnitude of decorrelations and consequently of rearrangement is less important than those in glass bead packing, probably due to the plastic deformation or/and creep flow at the PMMA particles contact. Moreover, the correlation function $max[\Gamma_{i,j}]$ tends to unity at the sintering step II and remains almost constant during the cooling step III. As expected, the rearrangement shall be considerably reduced in the sintered state of compacts. The similar evolution of correlation function is found with different sintering temperatures as shown in Fig. 8, supporting thus this interpretation.
Note however the correlation technique of acoustic speckles is no longer meaningful in the cooling step for the ultrasound transmission in sintered porous materials is dominated by coherent waves. To enhance scattered acoustic speckles, we should increase the working frequency of the wave which is however precluded by the high viscous dissipation of the PMMA material.
\begin{figure}[]
    \includegraphics*[width=7.5cm] {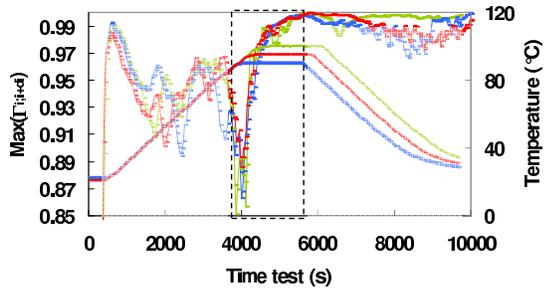}
  \caption{\label{fig:fig8} Cross-correlation function of high-frequency acoustic speckles at different temperatures of sintering for PMMA beads packings.
}
\end{figure}
\section{Conclusion}

	In summary, ultrasound transmission in PMMA bead packings during sintering has shown the transitional behaviour from the dominant transport of wave scattering in a noncohessive packing to coherent propagation in a cohesive compact. Velocity and amplitude measurements of the coherent longitudinal wave provide an efficient method to monitor the evolution of the elastic properties and the densification from the green body to sintered porous material. The comparison of our acoustic velocity data with the effective medium theory based on the Digby contact model reveals the crucial role of the bonding effect on the mechanical strength of the compacts during sintering, in conjunction with the increase of the contact area. In view of the difficulties in relating contact stiffness to measurable particle contact geometry during sintering [7, 15], we believe that the stiffnesses themselves ($D_n$ and $D_t$) are probably the fundamental parameters to characterize the interparticle interaction. The stiffnesses are less model dependant than contact sizes \cite{winkler83}; evaluating the stiffnesses from measured acoustic velocities would allow to monitor the evolution of the mechanical strength of powder compacts during sintering. 
Also, we have found that the amplitude measurement is much more sensitive than that of velocity to monitor the evolution of the contact networks of particles. This experimental finding shows that amplitude measurement may provide the useful information about the microstructure on smaller length scales of the coherent wavelength. Clearly, further careful measurements of attenuation are needed to understand the interplay of the elastic enhancement and the homogenisation of the contact networks in the course of sintering. With the sensitive probing by the correlation acoustic speckles, we have detected the important rearrangements at the initial stage of sintering which could be compared with numerical simulations in the future.
%\newpage %Just because of unusual number of tables stacked at end
\bibliography{sintering}% Produces the bibliography via BibTeX.

\end{document}